\documentclass[aps,pre,floats,twocolumn,superscriptaddress]{revtex4}
\usepackage{amssymb}

\usepackage{graphicx}
\usepackage[tbtags]{amsmath}
\usepackage{fancybox}
\usepackage{array}
\usepackage{color}
\usepackage{dcolumn}
\usepackage{amsmath}

\begin{document}
\title{Ensemble order parameter equations for star network}
\author{Yuting Sun}
\affiliation{Department of Physics and the Beijing-Hong Kong-Singapore Joint Centre for Nonlinear and Complex Systems (Beijing), Beijing Normal University, Beijing 100875, China}
\author{Jian Gao}
\affiliation{Department of Physics and the Beijing-Hong Kong-Singapore Joint Centre for Nonlinear and Complex Systems (Beijing), Beijing Normal University, Beijing
100875, China}
\author{Can Xu}
\affiliation{Department of Physics and the Beijing-Hong Kong-Singapore Joint Centre for Nonlinear and Complex Systems (Beijing), Beijing Normal University, Beijing 100875, China}
\author{Xia Huang }
\affiliation{Department of Mathematics and Physics, North China Electric Power University,Beijing 102206, China}
\author{Zhigang Zheng}
\email[]{zgzheng@bnu.edu.cn}
\affiliation{Department of Physics and the Beijing-Hong Kong-Singapore Joint Centre for Nonlinear and Complex Systems (Beijing), Beijing Normal University, Beijing 100875, China}
\date{\today}
\begin{abstract}
The OA ansatz has attracted much attention recently, infinite-dimensional Kuramoto model could collapses to a two-dimensional system of order differential equations with it. In this paper, we propose the ensemble order parameter (EOP) equations to describe the dynamics for networks with a finite size. To verify the effectiveness of this method, we apply it into the star network and star-connected network. In the star network, numerous phase transitions among different synchronous states are observed, three processes of synchronization, one process of de-synchronization and a group of hybrid phase transitions, the processes of those transitions are revealed by the EOP dynamics and other nolinear tools such as time reversibility analysis and linear stability analysis. Also in the star-connected network, the two-step synchronization transition is observed. The process of it is still be revealed by the similar methods in the single star network.

\end{abstract}
\maketitle
\section{Introduction}
Understanding the intrinsic microscopic mechanism embedding in collective macroscopic behaviors of populations of coupled units in heterogenous network has become a focus in a variety of fields, such an biological neurons circadian rhythm, chemical reacting cells, and even society systems~\cite{kuramoto1984chemical,acebron2005kuramoto,strogatz2000kuramoto,pikovsky2002synchronization,dorogovtsev2008critical,arenas2008synchronization,zheng1998phase}. Numerous different emerging macroscopic states/phases have been revealed, and various non-equilibrium transitions among these states have been observed and studied in heterogenous networks.

The phase transitions among different collective states in heterogenous network exhibit the typical feature of multistability, i.e., they may coexist for given parameters and depend on the choice of initial conditions. This interesting behavior is closely related to the first-order transition, especially the recently concerned explosive synchronization. Multistability and discontinuous transitions indicate the competitions of miscellaneous attractors and their corresponding basins of attraction in phase space. For a network of coupled oscillators, the miscroscopic discription of  the dynamics of oscillators should be made in a high-dimensional phase space, which is very difficult to deal with. The key point in understanding macroscopic transitions is the projection of the dynamics from this high-dimensional space to a much lower-dimensional subspace. This can be executed by introducing appropriate order parameters and building their dynamics. Especially the significant work of ~\cite{ott2008low}, in that case the original infinite-dimensional problem were reduced exactly to a flow in OA manifold which was described by several coupled ordinary differential equations(ODE). Moreover, for the case with finite oscillators, an ensemble method based on the OA manifold is also developed. Interestingly, the time-asymptotic attracting behavior of the full system and the attractor bifurcations are contained in this reduction and were quantitatively described by the ODE formulation.


Recently the abrupt transition from incoherent state to synchronization in network which the frequencies of oscillators on nodes are positively correlated to the node's degrees has attracted a lot of attention~\cite{gomez2011explosive}. This explosive transition to synchronization has been observed numerically on scale-free networks and experimentally in electronic circuits~\cite{leyva2012explosive,hu2014exact}. Moreover, it was also found that the first-order transition can be changed and numerous phase transitions can be observed by controlling the phase shift~\cite{xucan2015star}. Numerous efforts have been made to understand the mechanism of explosive synchronization from different viewpoints such as the topological structures of networks, the coupling functions among nodes, and so on~\cite{peron2012explosive,ji2013cluster,leyva2012explosive,zhang2013explosive,li2013reexamination,leyva2013explosive}. Although there have been numerous discussions and even analytical solution on understanding the explosive synchronization, it is still difficult to get an analytical insight in a high-dimensional phase space composed of dynamical variables of oscillators, and the dynamical analysis of transitions among different synchrony states is still lacking.

In this paper it is our mission to study the phase transitions among oscillators in a star network with the Sakaguchi-Kuramoto couplings by considering the effect of the phase shift among coupled oscillators. The star topology is the main topological characteristics of many heterogenous networks, e.g., the scale-free networks. The dynamics of star networks of oscillators is analytically studied by proposing the ensemble order parameter (EOP) and building the equations of motion of EOP for networks with a finite size,, which accomplishes a great reduction from microscopic high-dimensional phase dynamics of coupled oscillators to a macroscopic low-dimensional EOP dynamics. Based on the EOP dynamics, we further reveal, by using tools of nonlinear dynamics such as time reversibility analysis and linear stability analysis, that numerous phase transitions among different synchronous states in this model. We found three typical processes of the transitions to the synchronous state,
i.e., the transitions from the neutral state, the in phase state or the splay state to the synchronous state, and the continuous process of de-synchronization together with a group of hybrid phase transitions which are discontinuous with no hysteresis.
Furthermore, we study the synchronization dynamics of two coupled star networks and find a two-step synchronization transition, which can be macroscopically resorted to the dynamics of coupled EOP equations.

The present paper is organized as follows. In Section II, we illustrate the Ott-Antonsen ansatz and the EOP equations for network with a finite size.
In Section III, we introduce the Sakaguchi-Kuramoto model with star network connections and apply the EOP approach to project the dynamics of coupled oscillators to
the phase space of EOP. In Section \uppercase\expandafter{\romannumeral4}, the properties and mechanics of the phase transitions are analyzed in the plane of ensemble of order parameter.
Moreover, in the last section, the star-connected network is also discussed, and a two-step synchronization transition is found and analyzed with similar methods
 for the single star network. We believe that this is a meaningful step to understand the synchronization transition in heterogeneous networks.

\section{Ott-Antonsen ansatz}
Recently the issue of low-dimensional dynamics was reopened in work of ~\cite{ott2008low} with the discovery of an ansatz that collapses the infinite-dimensional Kuramoto model to a two-dimensional systems of ODEs.

To illustrate the ansatz briefly, let us apply it to the class of identical oscillators governed by the equation
\begin{equation}
\dot\varphi_{j}=fe^{i\varphi_{j}}+g+\bar{f}e^{-i\varphi_{j}},\quad j=1,\cdots,N,
\end{equation}
where $f$ is any smooth, complex-valued 2$\pi$-periodic function of the phases $\varphi_{1},\cdots,\varphi_{N}$ and g has to be real valued since $\dot\varphi_{j}$ is real. In the limit $N\longrightarrow\infty$, the evolution of the system (1) is given by the continuity equation
 \begin{equation}
\frac{\partial{\rho}}{\partial{t}}+\frac{\partial{\nu}}{\partial{\phi}}=0,
\end{equation}
where the phase density $\rho(\phi,t)$ is defined such that $\rho(\phi,t)d\phi$ gives the fraction of phases that lie between $\phi$ and $\phi+d{\phi}$ at time t, and where the velocity field is the Eulerian version of Eq.(1),
\begin{equation}
\nu(\phi,t)=fe^{i\varphi_{j}}+g+\bar{f}e^{-i\varphi_{j}}.
\end{equation}

 In the work of ~\cite{ott2008low}, suppose $\rho$ is of the form
 \begin{equation}
\rho(\phi,t)=\frac{1}{2\pi}\{1+\sum_{n=1}^{\infty}({\bar{\alpha}(t)^n}{e^{in\phi}}+\alpha(t)^n{e^{-in\phi}})\}
\end{equation}
 for some unknown function $\alpha$ that is independent of $\phi$. Note that Eq.(4) is just an algebraic rearrangement of the usual form for the Poissan kernel
 \begin{equation}
\rho(\phi)=\frac{1}{2\pi}\frac{1-r^2}{1-2r\cos(\phi-\Phi)+r^2}
\end{equation}
where r and $\Phi$ are defined via
 \begin{equation}
\alpha=r{e^{i\Phi}}.
\end{equation}
In geometrical terms, the ansatz (4) defines a submanifold in the infinite-dimensional space of density functions $\rho$. This Poisson submanifold is two dimensional and is parameterized by the complex number $\alpha$, or equivalently, by the polar coordinates r and $\Phi$.

The intriguing face discovered in work ~\cite{ott2008low} is that the Poisson submanifold is invariant: If the density is initially a Poisson kernel, it remains a Poisson kernel for all the time. To verify this, we substitute the velocity field (3) and the ansatz (4) into the continuity equation (2), and find that the amplitude equations for each harmonic $e^{in\phi}$ are simultaneously satisfied if and only if $\alpha(t)$ evolves according to
\begin{equation}
\dot\alpha=i(f{\alpha}^2+g\alpha+\bar{f}).
\end{equation}

This equation can be recast in a more physically meaningful form in terms of the complex order parameter denoted by $<z>$ and defined as the centroid of the phases $\phi$ regard as points $e^{i\phi}$ on the unit circle:
\begin{equation}
<z>=\int_0^{2\pi}{e^{i\phi}\rho(\phi,t)}d{\phi}.
\end{equation}
By substitute Eq.(4) into Eq.(8) we find that $<z>=\alpha$ for all states on the Poissan submanifold. Hence, <z> satifies the Riccati equation
\begin{equation}
\frac{d}{dt}<z>=i(f<z>^2+g<z>+\bar{f}).
\end{equation}

When $f$ and $g$ are functions of $<z>$ alone, as in mean-field models, Eq.(9) constitute a closed two-dimensional system for the flow on the Poissan submanifold.

For the finite K, we can get the similar property of the system through an ensemble way. By choosing an ensemble consisting of systems with the same parameter, we define an ensemble order parameter as
 \begin{equation}
\langle \alpha(t)\rangle=\langle\frac{1}{K}\sum_{j=1}^{K}e^{i\varphi_{j}}\rangle.
\end{equation}
where $<\cdot>$ means ensemble average, $K$ is the number of oscillators in a single system and $\varphi_{j},j=1,\dots,K$ are the phase variables for oscillators. With the approximated OA ansatz for network with finite size, if the initial phases of oscillators are chosen randomly from a Poisson kernel, we have the dynamical equation for the ensemble order parameter $\langle \alpha(t)\rangle$ as
\begin{equation}\label{equ:1}
\frac{d\langle\alpha(t)\rangle}{dt}=i(f\langle \alpha(t)\rangle^2+g\langle \alpha(t)\rangle+\bar{f}),
\end{equation}
with assumptions $\langle f\alpha(t)^{2}\rangle\approx f\langle\alpha(t)\rangle^2$ and $\langle g\alpha(t)\rangle\approx g\langle\alpha(t)\rangle$. With the assumptions work, this equation for ensemble order parameter can be used to get the low-dimensional collective behaviors of the system with finite size.

\section{STAR NETWORK WITH PHASE SHIFT}

In heterogeneous networks, e.g., the scale-free networks, nodes exhibit essentially different linking property, among which some nodes with dominantly large degree, i.e., hubs, play a dominate role. The simplest case of heterogeneous networks is the star topology, where a hub connects with other leaves with unit degree, can grasp the essential properties of heterogenity. As our starting point of discussion, we take a star-network of coupled phase oscillators with phase shift as our working model. The effect of phase shift among coupled oscillators has been extensively investigated in recent years, while this has been seldom discussed in star networks ~\cite{sakaguchi1986soluble,omel2012nonuniversal}.

Consider coupled phase oscillators with a star connecting topology. The star network is composed of a central node (i.e., the hub) and $K$ leaf nodes. Each of the leaf nodes connects solely to the hub. Thus, the degree of the leaves is $k_i=1$ ($i=1,...,K$) while that of the hub is $k_h=K$. Suppose that the natural frequency of the oscillators are proportional to their degree, the equations of motion for the hub and leaf nodes read
\begin{equation}
\begin{aligned}
\dot{\theta}_{h}&=\omega_{h}+\lambda\, \sum_{j=1}^{K}\sin\,(\theta_{j}-\theta_{h}-\alpha),\\
\dot{\theta}_{j}&=\omega+\lambda\,\sin\,(\theta_{h}-\theta_{j}-\alpha), \quad for \; 1\leq j\leq K,
\end{aligned}
\end{equation}
where $\theta_{h},\theta_{j}$ and $\omega_{h},\omega$ are instantaneous phases and natural frequencies of the hub and leaf nodes respectively. $\lambda$ is the coupling strength. $K$ is the number of leaf nodes connected with this hub and $\alpha$ is the phase shift. It was reported that abundant collective dynamics appear in the model with phase shift, such as synchrony can decay or incoherence can regain stability with increasing coupling and multistability between partially synchronized and/or the incoherent state can appear in the globally coupled network. The star network was original to study the characteristics of the explosive synchronization without phase shift, to explore the features of the transitions in the star network, it's necessary to investigate the system with phase shift. In such conditions, several new states are found, the characters and the stable conditions of them are analyzed in detail in the following part.

When $K=1$ the model is solvable. With the definition of phase difference $\varphi=\theta_{h}-\theta_{1}$ and difference of natural frequencies $\Delta\omega=\omega_{h}-\omega$, we have the dynamics for $\varphi$ as
\begin{equation}
\dot{\varphi}=\Delta\omega-2\lambda\sin(\varphi)\cos(\alpha)
\end{equation}
where $\cos(\alpha)$ is only an regulatory factor added to coupling strength $\lambda$. The synchronous state of the two oscillators means $\theta_{h}(t)=\theta_{1}(t)+c$ where $c$ is a constant, or the equivalent $\dot{\varphi}=0$. Hence the existence condition for synchronization reads
\begin{equation}
\lambda>\lambda_{c}=\frac{\triangle\omega}{2\cos(\alpha)}
\end{equation}
unless $\cos(\alpha)=0$. With $\lambda>\lambda_{c}$, the constant $c$ can be solved as $\varphi_{0}$ which satisfies the condition $\dot{\varphi}=0$, as $\varphi_{0}^{1}=\arcsin(\triangle\omega/(2\lambda\cos(\alpha)))$ or $\varphi_{0}^{2}=\pi-\varphi_{0}^{1}$. The two possible synchronous states is either stable or unstable which determined by $-\cos(\alpha)\cos(\varphi_{0})$ by the linear stability analysis. When $\cos(\alpha)>0$, the state $\varphi_{0}^{1}$ is stable and $\varphi_{0}^{2}$ unstable. When $\cos(\alpha)<0$, the case is just opposite. In the case of $\cos(\alpha)=0$, we have $\dot{\varphi}=\Delta\omega$ which is a constant and the system cannot be synchronous under any circumstances. Obviously, the phase shift $\alpha$ has great influence on the dynamics of the model in this case.

For the cases $K\geq2$, with phase differences $\varphi_{j}=\theta_{h}-\theta_{j}$, the star network with phase dynamics can be transformed to an all-connected network with phase difference dynamics as
\begin{equation}
\dot\varphi_{i}=\Delta\omega-\lambda\,\sum_{j=1}^{K}\,\sin(\varphi_{j}+\alpha)-\lambda\,\sin(\varphi_{i}-\alpha), 1\leq j\leq K
\end{equation}
To describe the synchronous state or non-synchronous state, we define the order parameter as
\begin{equation}
Z(t)\equiv r(t)e^{i\Phi(t)}=\frac{1}{K}\sum_{j=1}^{K}e^{i(\varphi_{j})}.
\end{equation}
It is instructive to rewrite Eq.(13) as
\begin{equation}
\dot\varphi_{j}=fe^{i\varphi_{j}}+g+\bar{f}e^{-i\varphi_{j}},\quad j=1,\cdots,K,
\end{equation}
where $i$ denotes the imaginary unit and $f=i\dfrac{\lambda}{2}$, $g=\Delta\omega-\lambda Kr\sin(\Phi+\alpha)$.

For the finite K, we can get property of the system through an ensemble way. By choosing an ensemble consisting of systems with the same parameter, we define an ensemble order parameter as
 \begin{equation}
z(t)\equiv \langle r(t)e^{i\Phi(t)}\rangle=\langle\frac{1}{K}\sum_{j=1}^{K}e^{i(\varphi_{j})}\rangle=\langle Z(t)\rangle.
\end{equation}
With the approximated OA ansatz ~\cite{ott2008low,marvel2009identical,marvel2009invariant} for network with finite size, if the initial phases of oscillators are chosen randomly in an fixed interval, we have the dynamical equation for the ensemble average of order parameter $z(t)$ as
\begin{equation}\label{equ:1}
\dot z=-\dfrac{\lambda}{2}e^{-i\alpha}z^{2}+i(\Delta \omega-\lambda Kr\sin(\Phi+\alpha))z+\dfrac{\lambda}{2}e^{i\alpha},
\end{equation}
and $Z(t)\approx z(t)$ from which we can use $z(t)$ to determine the state of the system instead of $Z(t)$. Eq.(8) describes the collective dynamics of Eq.(1) in terms of the order parameter, different solutions of Eq.(8) build correspondences with diverse collective states. Following we will analyze the characters and stable regions of those collective states with different parameters.

\subsection{Steady solutions}
The ensemble order parameter equation describes the collective dynamics of the star network. According to the Eq.(8), two steady solutions are found. The characters and the stable region in the phase diagram of them are analyzed as follows.

With Eq. (\ref{equ:1}), setting $z=x+iy$, we can describe the system in a x-y plane as
\begin{equation}
\begin{aligned}
\dot x&=\lambda(\dfrac{1}{2}+K)\cos\alpha\, y^{2}-\dfrac{\lambda}{2}\cos\alpha\, x^{2}\\
      &+\lambda(K-1)\sin\alpha\, x y-\Delta\omega\, y+\dfrac{\lambda}{2}\cos\alpha,\\
\dot y&=\lambda(\dfrac{1}{2}-K)\sin\alpha\, x^{2}-\dfrac{\lambda}{2}\sin\alpha\, y^{2}\\
      &-\lambda(K+1)\cos\alpha\, x y+\Delta\omega\, x+\dfrac{\lambda}{2}\sin\alpha,
\end{aligned}
\end{equation}
which can be studied analytically.

By setting $\dot x=0$ and $\dot y=0$, we have the four fixed points noted by $(x_{i},y_{i})$ with
\begin{equation}\label{equ:01}
\begin{aligned}
&x_{1,2}=\frac{-\sin\alpha\Delta\omega\pm A\sin\alpha}{\lambda(2K\cos2\alpha+1)}\\
&x_{3,4}=\frac{\sin\alpha}{\lambda}+\frac{\frac{\sin2\alpha}{2}B\pm K(\frac{\sin2\alpha}{2}B-\sin2\alpha^{2})}{\lambda\sin\alpha(K^{2}+2\cos(2\alpha)K+1)}\\
&y_{1,2}=-\frac{-\cos\alpha\Delta\omega\pm A\cos\alpha}{\lambda(2K\cos2\alpha+1)}\\
&y_{3,4}=\frac{-\Delta\omega(-\cos\alpha\pm\sin\alpha B-K\cos\alpha)}{\lambda(K^{2}+2\cos(2\alpha)K+1)}
\end{aligned}
\end{equation}
where $A=\sqrt{-2K\lambda^{2}\cos2\alpha-\lambda^{2}+\Delta\omega^{2}}$ and $B=\sqrt{\lambda^{2}+K^{2}\lambda^{2}+2K\lambda^{2}\cos2\alpha-\Delta\omega^{2}}$ which also give the existent condition for the fixed points as for point 1 and 2, as
\begin{equation}\label{equ:140}
\lambda\leq\lambda_{1}=\frac{\Delta\omega}{\sqrt{2K\cos2\alpha+1}},
\end{equation}
and for point 3 and 4, as
\begin{equation}
\lambda\geq\lambda_{2}=\frac{\Delta\omega}{\sqrt{K^{2}+2K\cos2\alpha+1}}.
\end{equation}
For point 3 and 4, there is an additional relation $x^{2}+y^{2}=1$ while for point 1 and 2, $x^{2}+y^{2}$ may be greater or lower than 1. For the definition of $z$, the point is allowable to describe the state of the system if and only if $x^{2}+y^{2}\leq1$.

As the analysis we did for the simple case with $K=1$, the linear stability analysis can be used for the fixed points, with which the Jacobian matrix is calculated,
\begin{equation*}
\small{
J=\left(\begin{matrix}
                   J_{11} & J_{12} \\
                   J_{11} & J_{21}
                   \end{matrix}
              \right)}
\end{equation*}
with $J_{11}=-\lambda\cos\alpha x+\lambda(K-1)\sin\alpha y$, $J_{12}=\lambda(1+2K)\cos\alpha y+\lambda(K-1)\sin\alpha x-\triangle\omega$, $J_{21}=\lambda(1-2K)\sin\alpha x-\lambda(K+1)\cos\alpha y+\triangle\omega$, $J_{22}=-\lambda\sin\alpha y-\lambda(K+1)\cos\alpha x$ ,the eigenvalues are
\begin{equation}
\beta_{1,2}=\frac{J_{11}+J_{22}\pm\sqrt{(J_{11}+J_{22})^{2}-4(J_{11}J_{22}-J_{12}J_{21})}}{2}.
\end{equation}
Then the stability of the fixed points can be judged from the eigenvalues of the matrix. If and only if the eigenvalues both are negative ,the fixed point is stable.

For point 1,the stability condition reads as
\begin{equation}
\lambda<\lambda_{c}^{f}=\footnotesize{\dfrac{\Delta\omega}{\sqrt{2K\cos2\alpha+1}}},\footnotesize{for\, \alpha \in(\alpha_{0}^{-}, 0)} \\
\end{equation}
where $\alpha_{0}^{-}=-\arccos(-1/K)/2$ and when $-\pi/2<\alpha\leq\alpha_{0}^{-}$ this point is always stable. It is worthy to note that the stability condition for it is the same with existence condition in Eq. (\ref{equ:140}) for $\alpha<0$. As for $\alpha>0$ it is always unstable. In the critical case with $\alpha=0$, the point has the neutral stability.
\begin{figure}
  \includegraphics[width=9cm,height=7cm]{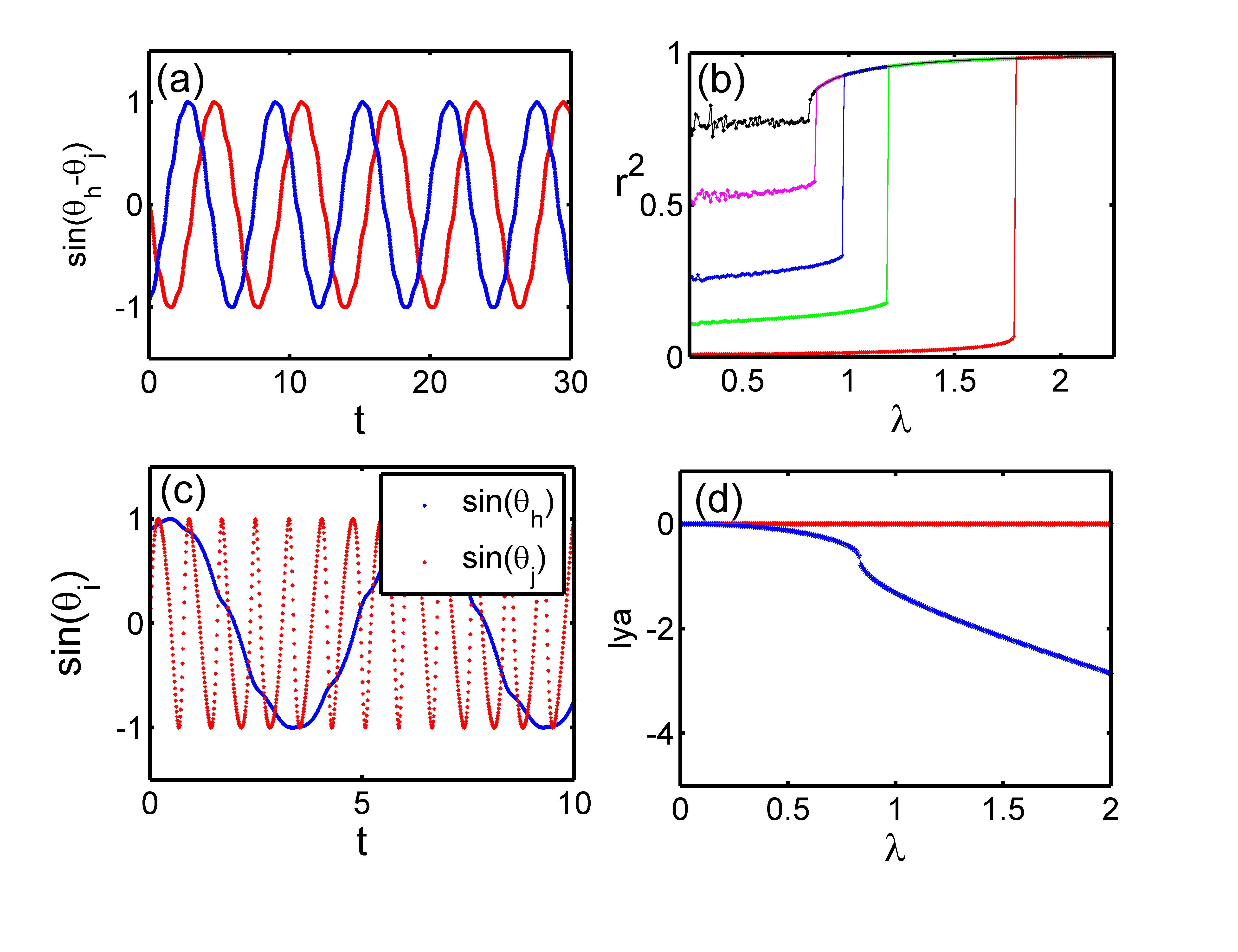}\\
  \caption{(color online)  ($a$) The time evolution of sin$\varphi_{j}$(t) with $\alpha=-0.4\pi,\lambda=2$, ($b$) The order parameter against the coupling strength with different initial states for $\alpha=0.0$, ($c$) The time evolution of sin$\theta_{i}$(t) with $\alpha=0.1\pi,\lambda=0.5$, ($d$) The lyapunov exponents of the network with $\alpha=0.1\pi$. The size of the star network is $N=11$.}
  \label{fig:1}
\end{figure}

For point 2, the stability condition reads as
\begin{equation}
\lambda>\lambda_{sc}^{+}=\footnotesize{\dfrac{-\Delta\omega}{K\cos2\alpha+1}}, \footnotesize{for\, \alpha \in(\alpha_{0}^{+}, \pi/2 )},
\end{equation}
where $\alpha_{0}^{+}=\arccos(-1/K)/2$.

For point 3,the stability condition reads as
\begin{equation}
\begin{cases}
\begin{array}{ll}
\lambda>\lambda_{sc}^{-}=\footnotesize{\dfrac{\Delta\omega}{K\cos2\alpha+1}}, & \footnotesize{for\, \alpha \in(\alpha_{0}^{-}, 0)}, \\
\lambda<\lambda_{sc}^{+}=\footnotesize{\dfrac{-\Delta\omega}{K\cos2\alpha+1}}, & \footnotesize{for\, \alpha \in(\alpha_{0}^{+}, \pi/2 )},
\end{array}
\end{cases}
\end{equation}
and when $0\leq\alpha\leq\alpha_{0}^{+}$ the point is always stable. It is worthy to note that, the critical coupling $\lambda_{sc}^{+}$ is special that when $\lambda>\lambda_{sc}^{+}$ the point 2 is stable while point 3 is unstable, and when $\lambda<\lambda_{sc}^{+}$ the case is just opposite. At last, for point 4, it is always unstable under any circumstances.

The fixed points of order parameter correspond to the stable solutions of coupled oscillators. For points 3 and 4, with $|z|=\sqrt{x^{2}+y^{2}}=1$ these two points correspond to the synchronous state (SS) of the system which means all the phase differences between hub and leaf nodes are the same and fixed as $\varphi_{j}(t)=constant$ for $1\leq j\leq K$. Hence the point 4 is always the unstable synchronous state and point 3 is the stable synchronous state in some special regions as we showed.

For points 1 and 2, if $|z|=\sqrt{x^{2}+y^{2}}>1$ then these states is not attainable because the order parameter $z$ is bounded by $|z|\leq 1$. If $|z|<1$ then the points are related to the splay state (SPS) which means all the phase differences satisfy a single function as $\varphi_{j}(t)=\varphi(t+\frac{jT}{K})$ with $T$ the period of $\varphi(t)$ for $1\leq j\leq K$, as shown in Fig. \ref{fig:1}(a). For $-1/2\pi<\alpha<0$, the point 1 is the stable splay state and point 2 is the unstable one in same regions and for $0<\alpha<1/2\pi$ the case is opposite, the stable splay state corresponds to the point 2.

\begin{figure}
  \includegraphics[width=9cm,height=6cm]{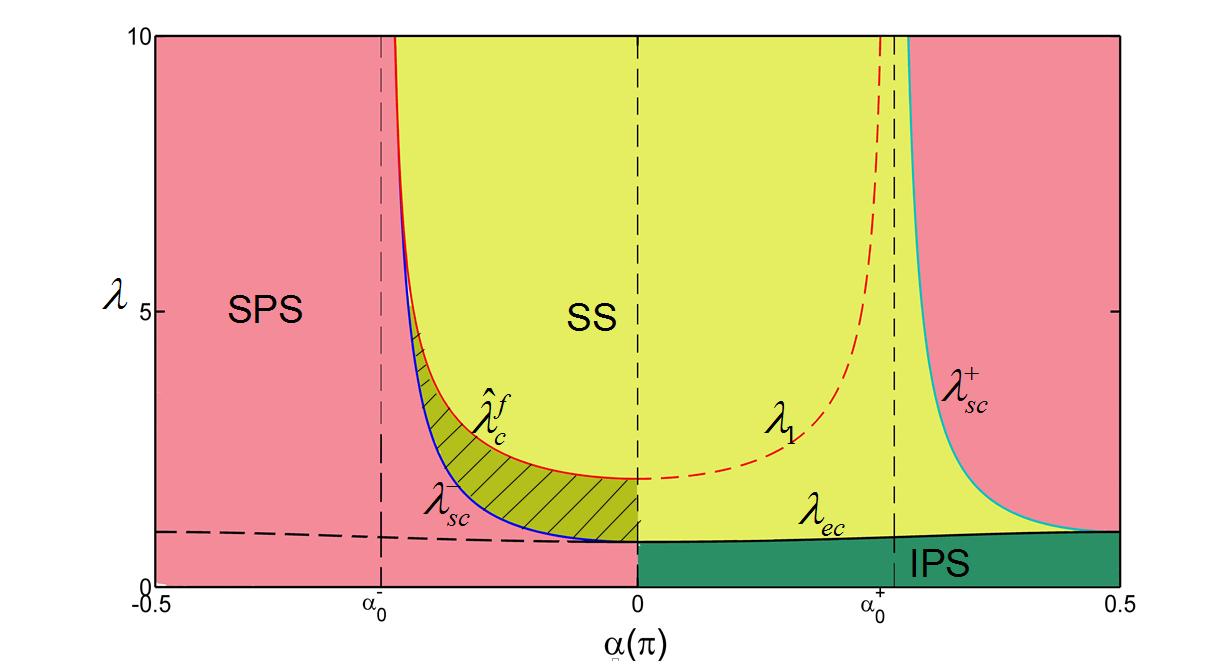}\\
  \caption{(color online) Phase diagram of system. Regime SS is stable synchronization region. Regimes SPS and IPS are stable region for splay state and in phase state respectively. The stable region for neutral state is too narrow to plot with only $\alpha=0,\pm\frac{\pi}{2}$. The coexistence regime of splay state and synchronous state is plotted by shadow.}
  \label{fig:2}
\end{figure}

All the existent and stable region for the synchronous state and splay state consist the phase diagram plotted in Fig. 2. It can be described by the four regions of the phase shift $\alpha$. When $-1/2\pi<\alpha<\alpha_{0}^{-}$, the splay state corresponding to point 1 always exist and is stable and the synchronous state exist if $\lambda>\lambda_{ec}=\lambda_{2}$ but always unstable. When $\alpha_{0}^{-}<\alpha<0$, the splay state exist and is stable with $\lambda<\hat\lambda_{c}^{f}=\lambda_{1}$, and the synchronous state exist with $\lambda>\lambda_{ec}$ but unstable unless $\lambda>\lambda_{sc}^{-}>\lambda_{ec}$. It is noted that there is a coexist region for the splay state and synchronous state when $\lambda_{sc}^{-}<\lambda<\hat\lambda_{c}^{f}$. In the third region, when $0<\alpha<\alpha_{0}^{+}$, the splay state is always unstable which only exist for $\lambda<\lambda_{1}$, and the synchronous state exist and is stable for $\lambda>\lambda_{ec}$. At last, when $\alpha_{0}^{+}<\alpha<1/2\pi$, the splay state exist all the time but only stable when $\lambda>\lambda_{sc}^{+}$, and the synchronous state only exist and stable in the region $\lambda_{ec}<\lambda<\lambda_{sc}^{+}$ in another hand.

\subsection{Periodical solutions}
With the analysis of fixed points, there are still two regions in the phase diagram where the stable state of system unclear, the one with $0<\alpha<1/2\pi$ and $\lambda<\lambda_{ec}$ where only exist unstable splay state, and the ones with $\alpha=0,\pm1/2\pi$ which is the critical cases. It is clearly that the periodical solutions must exist in those regions. Following, we will find the periodical solution and analyze the character and stable region of them.

For the first case, when $\lambda<\lambda_{ec}$ and $0<\alpha<\pi/2$, it is shown that there is no stable fixed point exist. However, from the numerical simulation and lyapunov exponents for the trajectories, shown in Fig. 1(d), we find the largest lyapunov exponent is zero and the others are negative which means there exist a limit cycle as the stable state and the volume of phase space is convergent.

To find the limit cycle, we transfer the Eq .(8) from x-y plane to polar coordinates by setting $z=re^{i\Phi}$,
\begin{equation}
\begin{aligned}
&\dot r=-\dfrac{\lambda}{2}(r^2-1)\cos(\Phi+\alpha),\\
&\dot\Phi=-\dfrac{\lambda}{2}(r+\dfrac{1}{r})\sin(\Phi-\alpha)+\triangle\omega-\lambda Kr\sin(\Phi+\alpha).\\
\end{aligned}
\end{equation}
It is easy to see that there is a limit cycle solution with $r=1$ and periodic phase $\Phi(t)$ for $\lambda<\lambda_{ec}$. This state we call in phase state(IPS), where all the phase differences are the same as $\varphi_{j}=\varphi(t)$ for $1\leq j\leq K$. This state means in the dynamical model Eq.(1) all the phases of leaves nodes are identical, as shown in Fig. \ref{fig:1}(c). In this case, the model Eq.(1) is reduced to the case with $K=1$ and the stability of the state can be got by the Flouquet theory for limit cycle as it is stable for $0<\alpha<\pi/2$ and unstable for $-1/2\pi<\alpha<0$.
\begin{figure}
  \includegraphics[width=9cm,height=4cm]{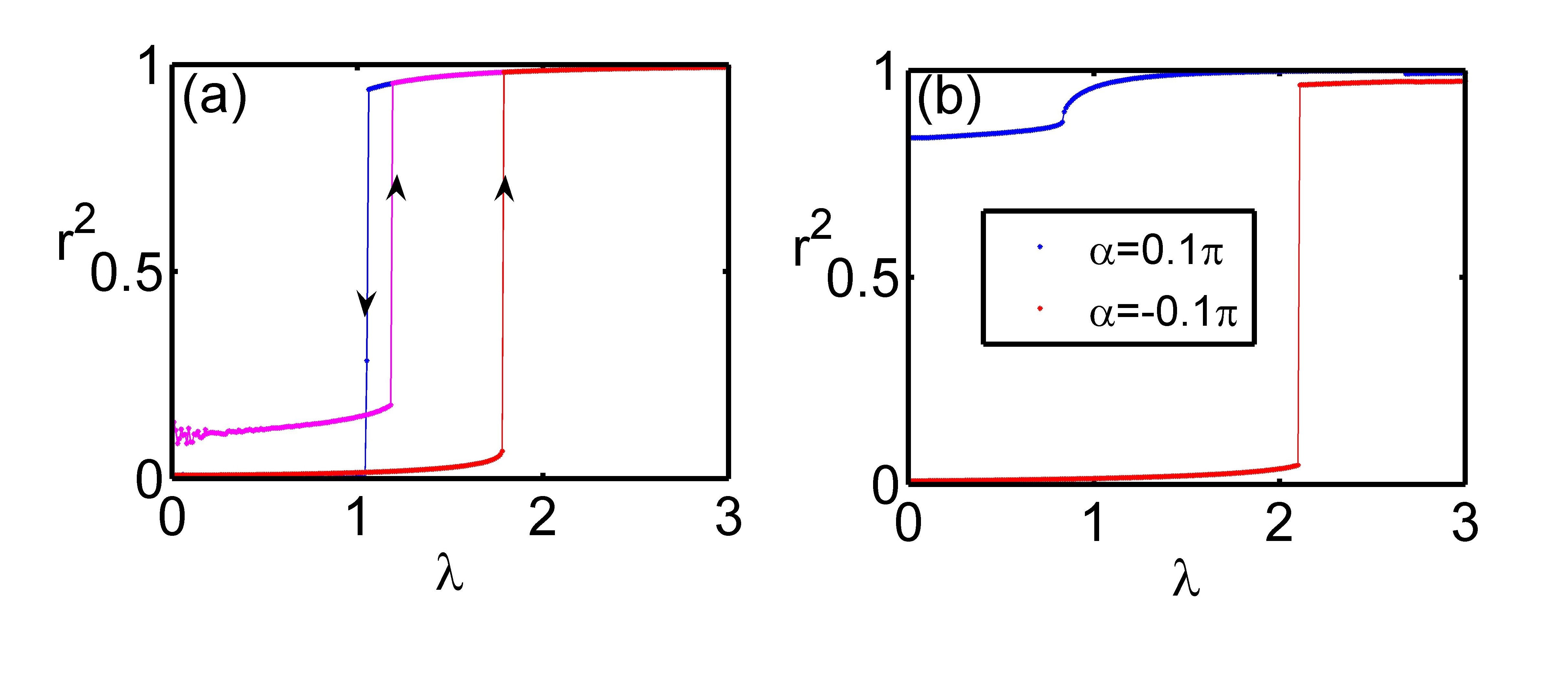}\\
  \caption{(color online) ($a$) The forward and backward continuation diagrams in different initial conditions with $\alpha=0$, ($b$) The forward continuation diagram with $\alpha=\pm0.1\pi$.}
  \label{fig:3}
\end{figure}

For the second case, when $\alpha=0,\pm1/2\pi$, these are the critical cases for the system. In the above analysis for $K=1$, with $\alpha=\pm1/2\pi$ we have the conclusion that the system will never be synchronous. Meanwhile, the fixed point 1 is found to be neutral stable when $\alpha=0,\pm1/2\pi$. With numerical simulation, we find that there is a large class of states in the critical cases with the order parameter can be almost every number between 0 and 1, determined by the initial situation of the nodes, shown in Fig. 1(b). When $\alpha=0$ there are transitions from the special state to synchronous state at different coupling strength. And when $\alpha=\pm1/2\pi$ the system will always in the states whatever the coupling strength is. Moreover, the sum of Lyapunov exponents for these states is zero which means the volume of phase space is conserved and hence we call them the neutral states (NS).

To understand the mechanics of these neutral states, in the framework of approximated Ott-Andeson ansatz and the plane of ensemble order parameter, we get the dynamical equation in with $\alpha=0$ as
\begin{equation}\label{equ:2}
\begin{aligned}
\dot x &=\lambda(K+\frac{1}{2})y^{2}-\frac{\lambda}{2}x^{2}-\Delta\omega y +\frac{\lambda}{2},\\
\dot y &=-\lambda(K+1)xy+\Delta\omega x.
\end{aligned}
\end{equation}
By setting $\dot x=0$ and $\dot y=0$,when $\lambda<\lambda_{ec}$ only one fixed point exist inside the unit cycle in the plane as the point 1 we analyzed.

According to the linear stability the fixed point is neutral stable. It is worthy to note that if we define an time reversal transformation as $R:(t,x,y)\mapsto(-t,-x,y)$, with $R$ effect on both sides of the equations, the dynamical equations remains invariant and hence they are called the time reversal dynamical system. This is a special kind of symmetry of dynamical system, which is also called the quasi-Hamiltonian~\cite{topaj2002reversibility}. This symmetry gives the system many interesting properties.
\begin{figure}
  \includegraphics[width=8.5cm,height=3.5cm]{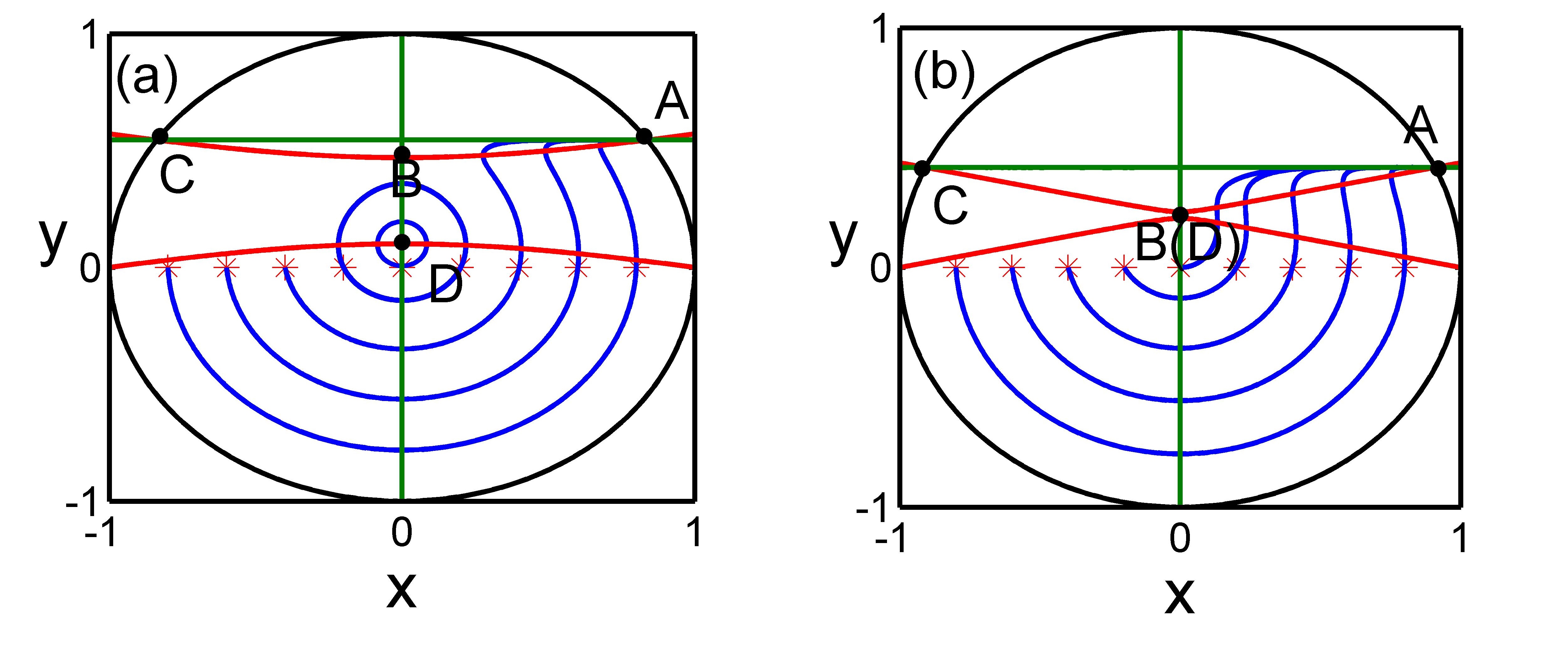}\\
  \caption{(color online) Phase plane of Eqs.(\ref{equ:2}) with $\Delta\omega=9$, $K=10$, $\alpha=0$ ($a$) with $\lambda=1.5$, ($b$) with $\lambda=1.9$. Red lines are $\dot x=0$, and green lines are $\dot y=0$. The intersections of $\dot x=0$ and $\dot y=0$ are fixed points A, B, C, D. Trajectories with different initial values are  marked by '$\ast$'.}
  \label{fig:4}
\end{figure}

Note that, the time reversal transformation $R$ can be resolved into $R=TW$ with $T:t\mapsto-t$ and $W:(x,y)\mapsto(-x,y)$. Hence the invariant set for $W$ is the $y$ axis with $x=0,y>0$. For any trajectory cross this invariant set, according to the time reversal symmetry, the forward trajectory and the backward trajectory are symmetrical. If the forward trajectory evolves to an attractor then there will be an symmetrical repeller of the system, which is the same attractor for backward trajectory. If the trajectory cross the invariant set more than once, then the forward and backward trajectory will coincide with each other, forming the periodic solution for the system, which is called reversible trajectory. For any reversible trajectory, the Lyapunov exponents have the sign-symmetry form and the volume of phase space in the vanity of it are conserved as we discovered with numerical simulation.

For our planer system bounded by unit circle with the invariant set as $x=0,y>0$, when the only fixed point is neutral stable, every trajectory is closed periodic one which belongs to the neutral states. This is what happens when $\alpha=\pm1/2\pi$ for all the coupling strength and $\alpha=0$ with the region $\lambda<\lambda_{ec}$. For $\alpha=0$ and $\lambda>\lambda_{ec}$, there is an coexist region for the synchronous state and neutral states as the critical cases for the coexist region for splay state and synchronous state.

Up to now, we get the phase diagram completely as shown in Fig .2. There are three main states, namely, the splay state, in phase state and synchronous state and one kind of critical state, the neutral state. The stable region for each state is plotted in the phase diagram with the boundaries we get analytically above both from existence and stability conditions. There is also a coexistence region for splay state and synchronous state, with the critical case for neutral state and synchronous state. The variety of states in the phase diagram leads to various phase transition of the system.

\subsection{Processes of synchronization and de-synchronization}
As we see in the phase diagram, the model we use has great variety of phase transitions among the four special states, as the transitions to synchronous state, namely synchronization, and correspondingly de-synchronization, also the transition between the same kind of states when crossing the critical state with $\alpha=0,\pm\frac{1}{2}\pi$. Some of the states have coexistent regions, which lead to abrupt transitions among them and hysteresis behaviors, while the others lead to transitions either abrupt or continuous, according to the kind of bifurcation at the critical point.

Among all the transitions, the processes of synchronization are most important. In the phase diagram it is clear that there are three processes to synchronous state from splay state, in phase state and neutral state respectively, with fixed phase shift $\alpha$ and increasing coupling strength $\lambda$ as usually assumed.

The first, we take an view on the synchronization process as neutral state to synchronous state with $\alpha=0$. The synchronization process is discontinuous known as the explosive synchronization which has attracted much attention recently. According to numerical computations this kind of transition is revealed abrupt, and there is a hysteretic behavior at the onset of synchronization. $\lambda_{c}^{b}$ and $\lambda_{c}^{f}$ are the backward and forward critical coupling strengths respectively, where $\lambda_{c}^{b}=\lambda_{2}$ and  $\lambda_{c}^{f}$ depends on initial states as shown in Fig. \ref{fig:3}(a). The upper limit of $\lambda_{c}^{f}$ is denoted by $\hat\lambda_{c}^{f}$. As $\lambda>\hat\lambda_{c}^{f}$, the synchronization state is globally attractive. Apparently it is difficult to understand the process of it according to the self consistent method, especially for the hysteresis behavior and coexistent region.
\begin{figure}
  \includegraphics[width=9cm,height=3.5cm]{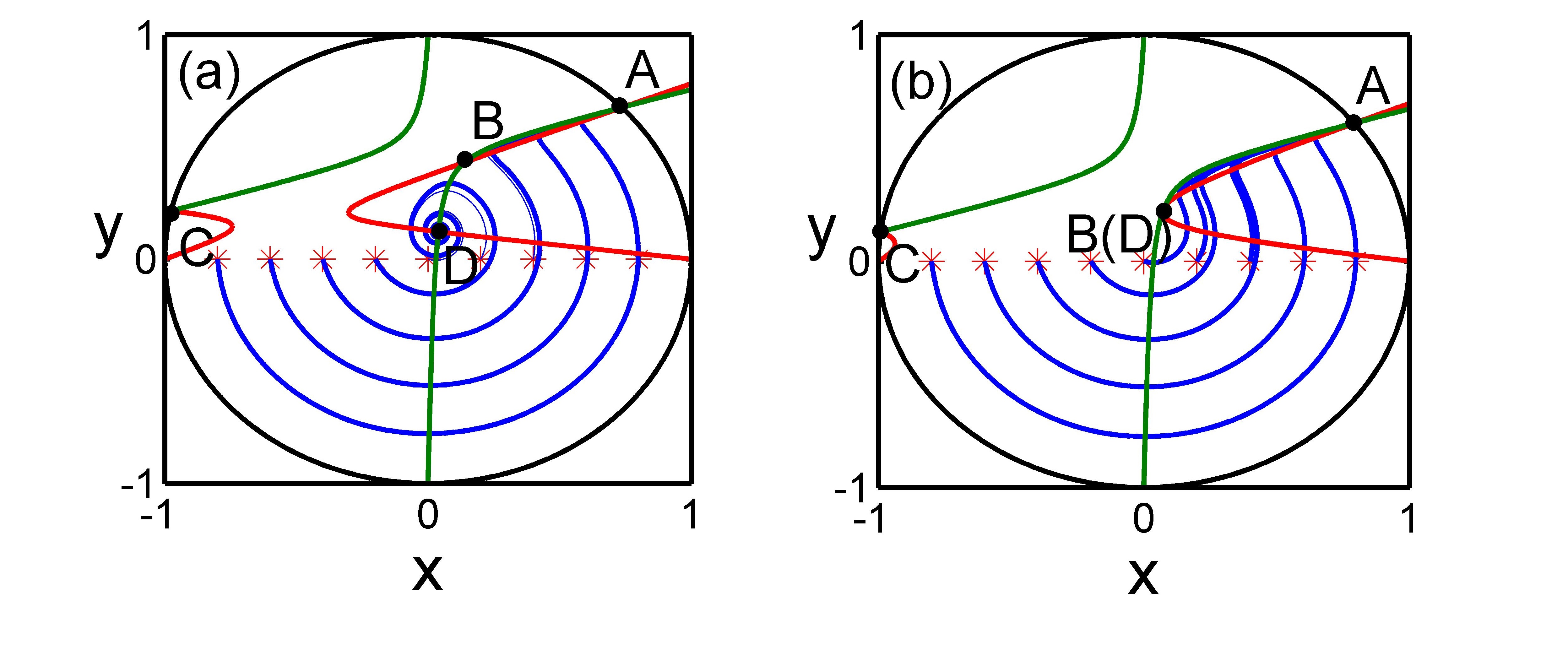}\\
  \caption{(color online) Phase plane for $\Delta\omega=9$, $K=10$, $\alpha=-0.1\pi$, ($a$) with $\lambda=1.8$, ($b$) with $\lambda=2.17$. Red lines are $\dot x=0$, and green lines are $\dot y=0$. The intersections of $\dot x=0$ and $\dot y=0$ are fixed points A,B,C,D. Trajectories with different initial values are marked as '$\ast$'. }\label{fig:5}
\end{figure}

However, as we see, through the method of OA anstaz, the coexistence of neutral state and synchronous state is pronounced, and the basins of attraction of each state can be determined also. For the time reversal symmetry the system has, we have defined an time reversal transformation as $R=TW:(t,x,y)\mapsto(-t,-x,y)$ and the invariant set for $W$ is $x=0,y>0$. In the case with the planer system bounded by unit circle, every trajectory either evolves to an stable point in the limit cycle or is closed periodic one with neutral stability. The separation between the two states is the condition for the trajectory to evolve only one round, as
\begin{equation}
\begin{aligned}
&\frac{dx}{dt}=0,\frac{d^{2}x}{dt^{2}}x>0\\
&\frac{dy}{dt}=0,\frac{d^{2}y}{dt^{2}}y>0
\end{aligned}
\end{equation}
The trajectories are plotted in Fig. \ref{fig:4}. The separation is the up red line with which the basin of attraction for neutral state and stable synchronous state are separated. The points D and B are the points 1 and 2 corresponding to the splay state and points A and C are the points 3 and 4 corresponding to the synchronous state. When $\alpha=0$, point D is neutral stable and point B is an saddle. Points A and C are linear stable and unstable respectively. The neutral states are the closed periodic orbits around the point D.

With the changing of parameters as coupling strength $\lambda$ or difference of natural frequency $\Delta\omega$ the separation line changes and phase transition will happens between these two steady states with a single initial state with marks $*$ in Fig. \ref{fig:4}. The transition is abrupt or explosive apparently, as shown in Fig. \ref{fig:3}(a), with hysteresis. And it is easy to take the basion for neutral state approximated by the circle which has its center in point $D$ and radius as the length of line $B-D$.

When $\alpha<0$ it can be shown that the point D becomes linear stable. If it is the only stable point in the phase space, all the trajectories will evolve to it as splay state of the system. In the coexistent region with another stable point A, which corresponding to the synchronous state, both the two stable points has their own basion of attraction. In this case, the point B plays the role of the separations of the attraction regions for both stable state. It is shown in Fig. \ref{fig:5}. With the parameter changing, the point B as the separation changes. The transition from splay state to synchronous state is also an abrupt one, shown in Fig. \ref{fig:3}(b), also with hysteresis.
\begin{figure}
  \includegraphics[width=9cm,height=4cm]{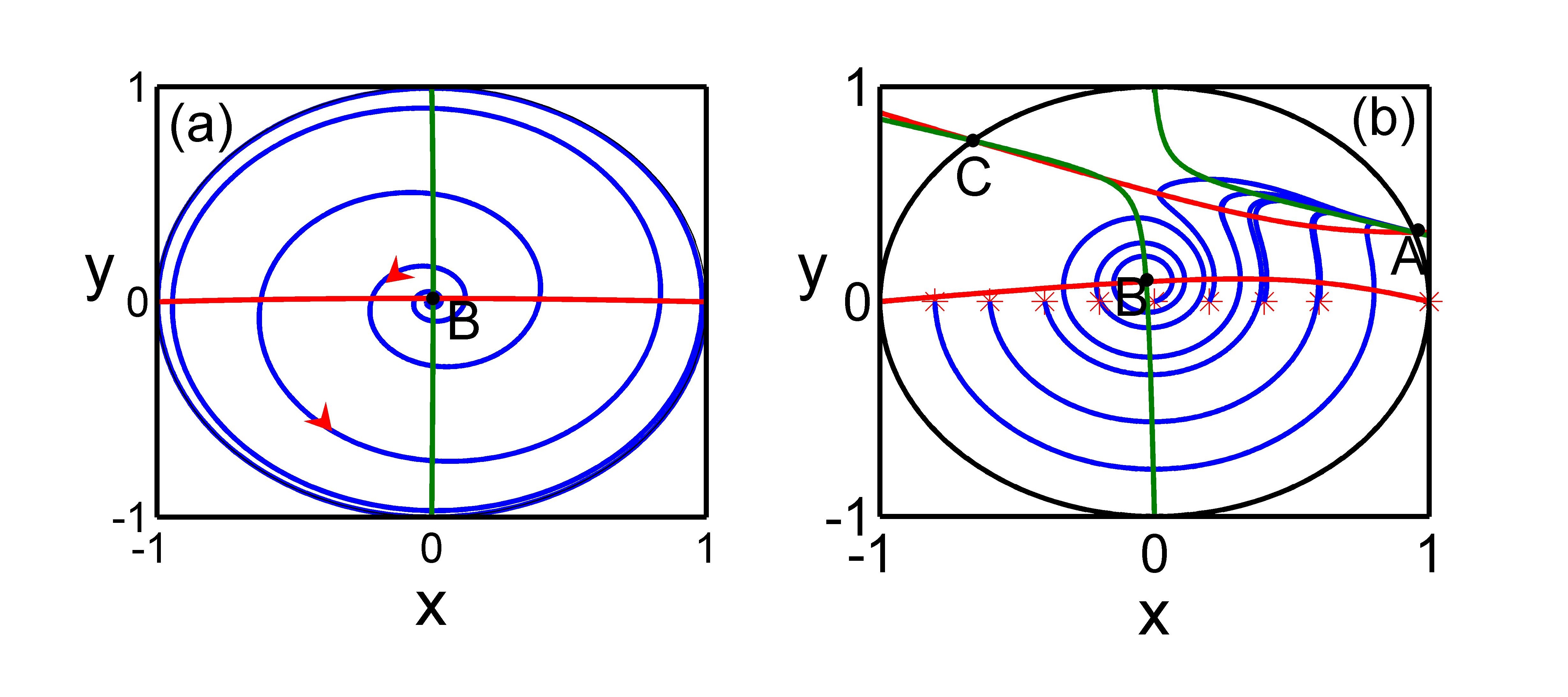}\\
  \caption{(color online) Phase plane for $\Delta\omega=9$, $K=10$, $\alpha=0.3\pi$, ($a$) with $\lambda=0.5$, ($b$) with $\lambda=1.5$. Red lines are $\dot x=0$, and green lines are $\dot y=0$. The intersections of $\dot x=0$ and $\dot y=0$ are fixed points A,B,C,D. Trajectories with different initial values are marked as '$\ast$'. }\label{fig:6}
\end{figure}

When $\alpha>0$ the points D and B are both unstable and before the appears of points A and C there is no stable points in the unit circle. The trajectories in this case will all evolves to the boundary which is a limit circle for the planar system. Hence when a stable point A and unstable point C appears in the unit circle the transition from the in phase state to synchronous state is continuous as the ghost effect of stable fixed points which is shown in Fig. \ref{fig:3}(b). The process of phase transition in the planer phase space is shown in Fig. \ref{fig:6}.

The three ways to synchronization discussed above are either abrupt or continuous. As we see, the different routes to synchronization is based on the different kinds of incoherent state, namely, the neutral state with $\alpha=0$, the splay state with $\alpha<0$ and the in phase state with $\alpha>0$. The small change of $\alpha$ around zero can change the process of synchronization dramatically. This is the reason we are interested in the model with phase shift.

\begin{figure}
  \includegraphics[width=9cm,height=7cm]{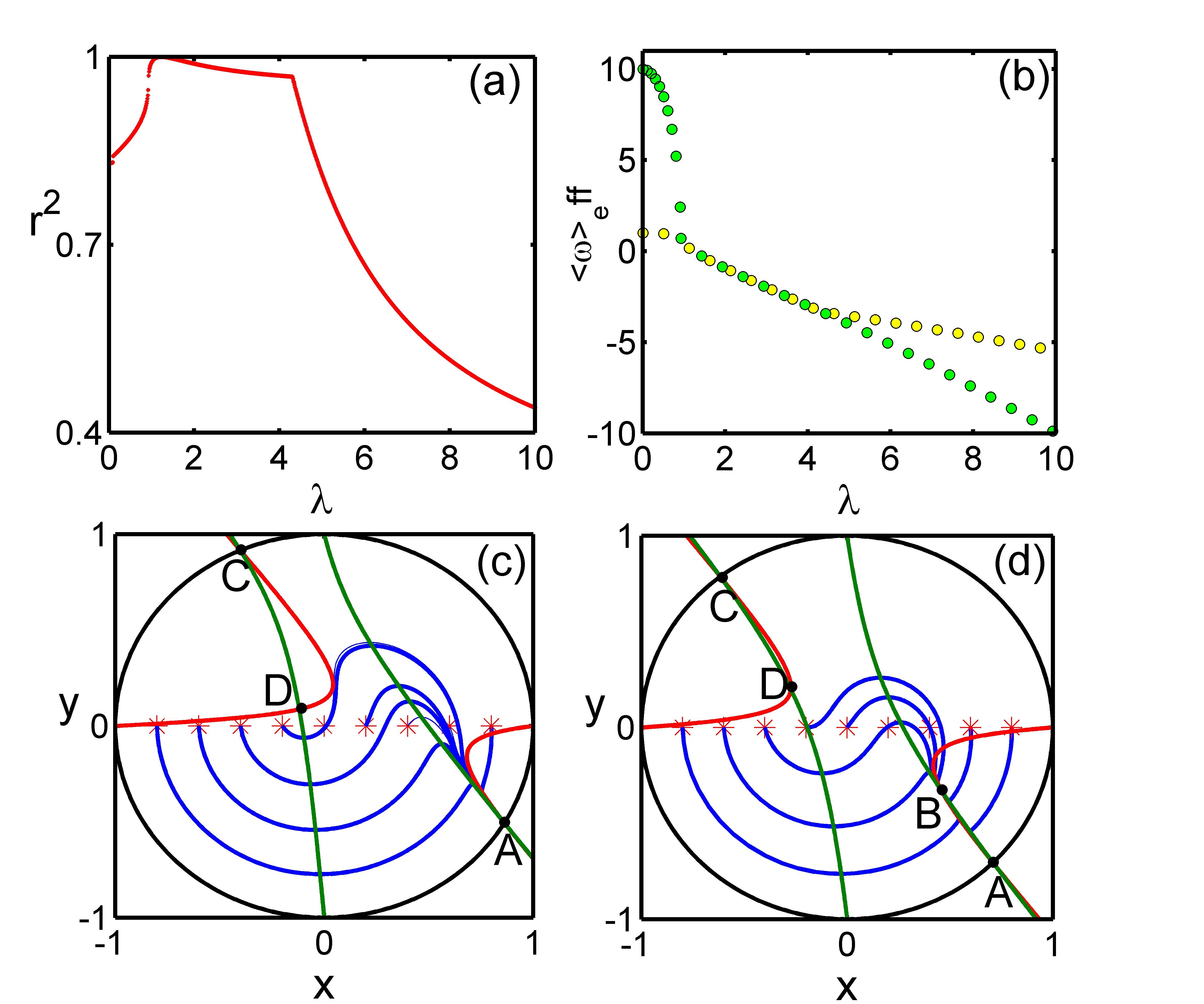}\\
  \caption{(color online) ($a$) The order parameter against the coupling strength with $\alpha=0.3\pi$. ($b$) The effective frequency of the hub nodes(the green) and the leaf nodes(the yellow) against the coupling strength with $\alpha=0.3\pi$. Phase plane for $\Delta\omega=9$, $K=10$, $\alpha=0.3\pi$, ($c$) with $\lambda=3$, ($d$) with $\lambda=5$. Red lines are $\dot x=0$, and green lines are $\dot y=0$. The intersections of $\dot x=0$ and $\dot y=0$ are fixed points A,B,C,D. Trajectories with different initial values are marked as '$\ast$'.}\label{fig:7}
\end{figure}

 Contrary to the conventional belief that the system will always be synchronous if we set the coupling strength large enough, the synchronous state is unstable when $\lambda>\lambda_{sc}^{+}$. It is clear in the phase diagram. The process is de-synchronization which is shown in Fig. \ref{fig:7}(a) and (b). The order parameter decrease from 1 and effective frequencies of hub and leaf nodes are divided at the same time. This a transition from synchronous state to splay state.

When $\lambda<\lambda_{sc}^{+}$, the system is in synchronous state, namely the point A in the phase plane. At the critical case $\lambda=\lambda_{sc}^{+}$, the red line of $\dot x=0$ and the black line of $\dot y=0$ are tangent in the unit cycle in fixed point A where the other fixed points D and C are still unstable. If we increase the coupling strength and let it larger than $\lambda_{sc}^{+}$, the two nullclines will intersect in two fixed points A and another one, the point B. The point A loses its stability and the new point B is a stable one corresponding to the splay state. The process from the synchronous state to splay state is finished by this bifurcation continuously. If we reverse this process by decreasing coupling strength, it is not doubt a process of synchronization, although we usually not see it in this view.

\subsection{Hybrid transition}

Up to now, we have discussed four kinds of phase transitions, either abrupt with hysteresis or continuous. All of them are concerned with synchronous state and we follow the tradition with fixed phase shift $\alpha$ and increasing coupling strength $\lambda$. However, there is still a class of phase transitions which seems to be interesting for us, with fixed coupling strength $\lambda$ and changing the phase shift $\alpha$, which allows us to think the process across the critical lines with $\alpha=0,\pm1/2\pi$.

\begin{figure}
  \includegraphics[width=9.5cm,height=8cm]{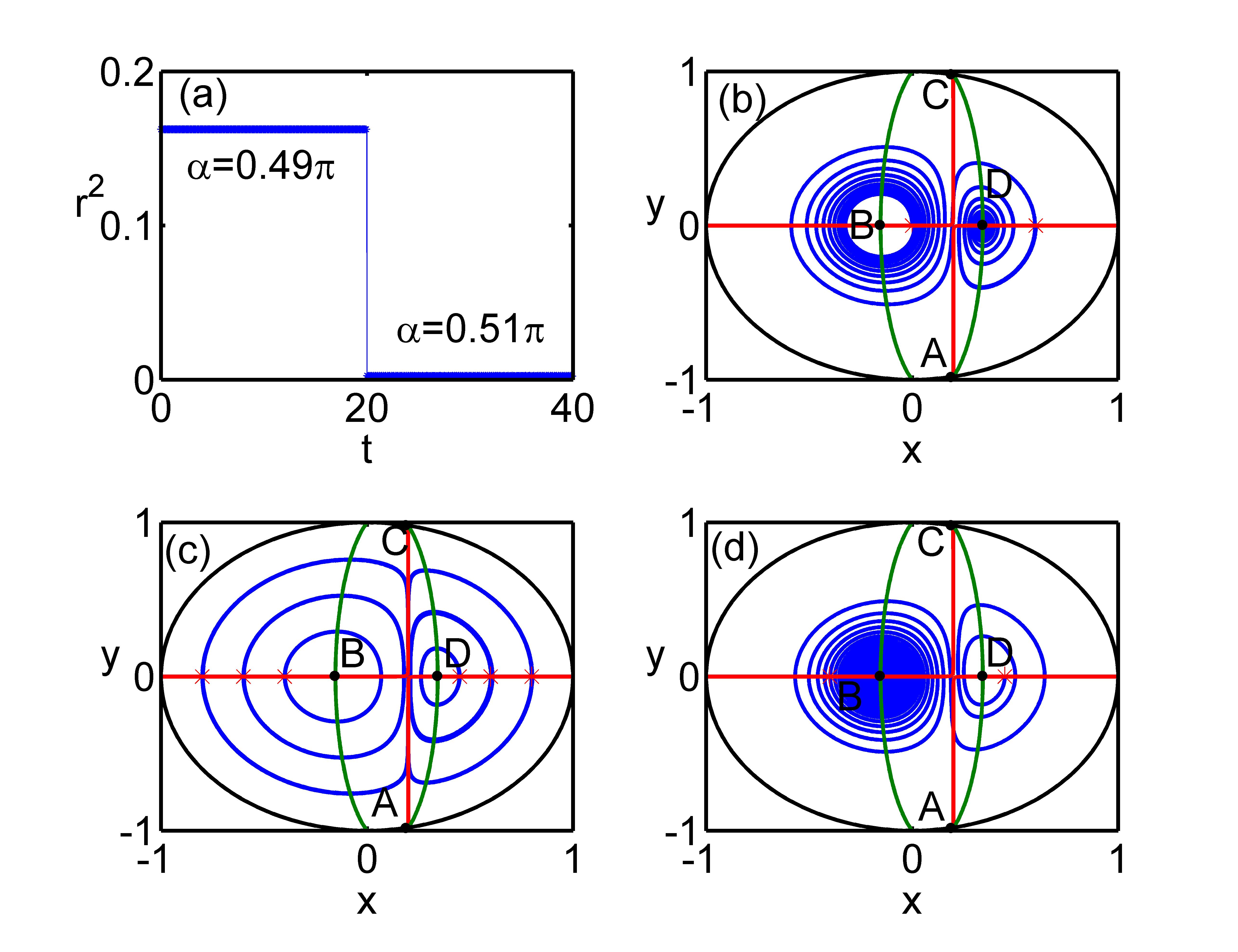}\\
  \caption{(color online) ($a$) The order parameter against time in different phase shift $\alpha$ with $\lambda=5$. Phase plane for $\Delta\omega=9$, $K=10$, $\lambda=5$, ($b$) with $\alpha=0.49\pi$, ($c$) with $\alpha=0.50\pi$, ($d$) with $\alpha=0.51\pi$. Red lines are $\dot x=0$, and green lines are $\dot y=0$. The intersections of $\dot{x}=0$ and $\dot{y}=0$ are fixed points A,B,C,D. Trajectories with different initial values are marked as '$\ast$'.}\label{fig:8}
\end{figure}

Surprisingly, at a glaze, this crossing process always leads to a kind of hybrid transition which is abrupt but without hysteresis, from splay state to in phase state, and even between synchronous states and splay states in the different sides of these lines, as shown in Fig.\ref{fig:8}(a) for transitions between splay states.

However the reasons of it is not very difficult to understand. Based on the analysis above for the cases with $\alpha=0,\pm1/2\pi$, they are protected by the time-reversal symmetry as the critical cases. Once the phase shift crossing the critical cases, all the stability of the solutions will turn to the opposite, which is typical for hybrid transition.

Take the phase transition between splay states as an example. In the phase plane of order parameter, there exist four fixed points for each side of the system. One stable point and one unstable point inside the unite cycle with two saddle points in it.  As shown in Fig.\ref{fig:8}(b), the point D is stable and B unstable. When crossing the critical case, both the points D and B are neutral stable in Fig.\ref{fig:8}(c). And in other side of the critical case, the stability of them turn to the opposite which means the saddle points are still the saddle points but the stable point becomes unstable and vice versa. It is shown in Fig.\ref{fig:8}(d) with the point B the stable one and D the unstable one. The process from (b)-(c)-(d) or (d)-(c)-(b) in Fig.\ref{fig:8} is a typical hybrid transition.

Noting that all the states are the collective behaviors of the phase oscillators in the star network, the phase transitions here are not the trivial ones at all. This is largely related to the symmetry of the system, both the identity of the oscillators which makes the approximate OA ansatz appropriate and the time-reversal symmetry for which phase shift affects mostly. Among all the cases, the most important one is the case with $\alpha=0$, which has the time-reversal symmetry and also the original model to be used for discussion of synchronization.

\section{STAR-CONNECTED NETWORK}
In the heterogeneous network especially scale-free network, hubs play the dominate roles. The single star network which get the essential properties of hubs is very useful to explore the main behavior of heterogeneous network which we have discussed above. However, the heterogeneous network such as scale-free network have several hubs, and hubs has influences on each other. To better describe the phase transitions in the heterogeneous network, we can combine several star networks together and let the hubs connected directly, namely the star-connected network.

\subsection{Model and the simplification}
Considering the simplest network composed by two star networks, namely star 1 and star 2. The model reads
\begin{equation}
\begin{aligned}
\dot{\theta}_{h_{1}}&=\omega_{h_{1}}+\lambda_{1}\, \sum_{j_{1}=1}^{K_{1}}\sin\,(\theta_{j_{1}}-\theta_{h_{1}})+{\lambda_0}\sin(\theta_{h_{2}}-\theta_{h_{1}}),\\
\dot{\theta}_{j_{1}}&=\omega_{1}+\lambda_{1}\,\sin\,(\theta_{h_{1}}-\theta_{j_{1}}), \quad for \; 1\leq j_{1}\leq K_{1},\\
\dot{\theta}_{h_{2}}&=\omega_{h_{2}}+\lambda_{2}\, \sum_{j_{2}=1}^{K_{2}}\sin\,(\theta_{j_{2}}-\theta_{h_{2}})+{\lambda_0}\sin(\theta_{h_{1}}-\theta_{h_{2}}),\\
\dot{\theta}_{j_{2}}&=\omega_{2}+\lambda_{2}\,\sin\,(\theta_{h_{2}}-\theta_{j_{2}}), \quad for \; 1\leq j_{2}\leq K_{2},
\end{aligned}
\end{equation}
where ${\theta}_{h_{1}},{\theta}_{h_{2}},{\theta}_{j_{1}},{\theta}_{j_{2}}$ and ${\omega}_{h_{1}},{\omega}_{h_{2}},{\omega}_{j_{1}},{\omega}_{j_{2}}$ are phases and natural frequencies of the hub and leaf nodes in star 1 and star 2 respectively. $K_{1},K_{2}$ and $\lambda_{1},\lambda_{2}$ are the numbers of leaf nodes and coupling strength for star 1 and 2. $\lambda_{0}$ is the coupling strength between the two stars.

\begin{figure}
  \includegraphics[width=9.5cm,height=8cm]{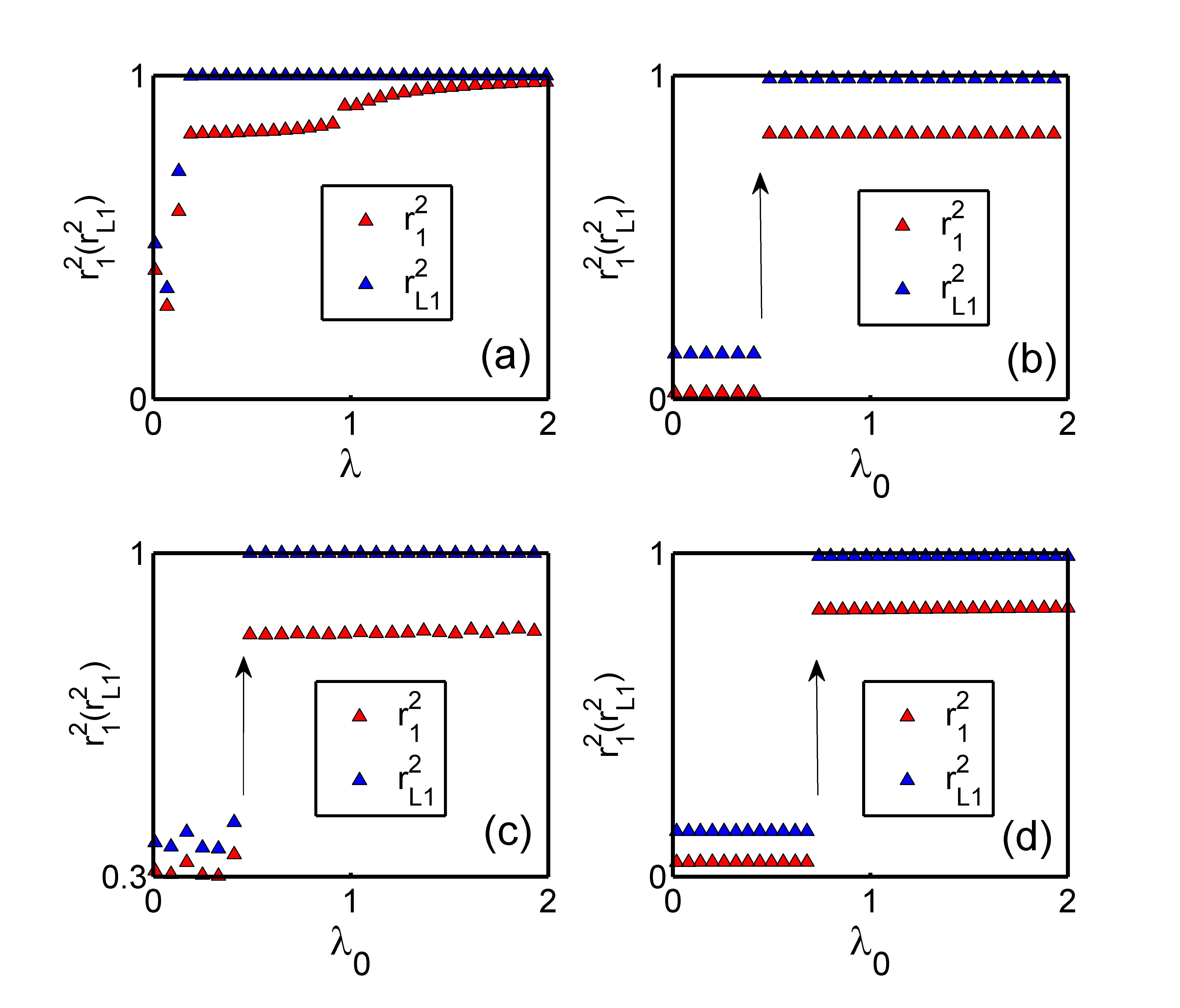}\\
  \caption{(color online) The order parameter of star 1 $r_1$ and the local order parameter of leaf nodes of star 1 $r_{L1}$ against the coupling strength $\lambda$ or $\lambda_0$ with different coupled forms when $K_1=10, K_2=9, \lambda_1=\lambda_2=\lambda$. ($a$) $\lambda_0=\lambda$, bidirectional coupled with the star 2. ($b$) $\lambda=1.3$, bidirectional coupled with the star 2. ($c$) $\lambda=1.3$, unidirectional coupled by the star 2, ($d$) $\lambda=1.3$, unidirectional coupled by the hub of star 2.}\label{fig:9}
\end{figure}

With $\alpha=0$, in the single star network without phase shift, the phase transition is from neutral state to synchronous state with abrupt transition. However, when coupled together, setting $\lambda_{1}=\lambda_{2}=\lambda_{0}\equiv\lambda$, the synchronization process with increasing $\lambda$ for a single star network is divided into two parts. A intermediate state is observed that the leaf nodes of each star will synchronize before the synchronous state appears. The process appears to be from neutral state to in phase state abruptly then to synchronous state continuously, if we borrow the definition for the single star network, as shown in Fig.\ref{fig:9}(a).

The second step of transition from in phase state to synchronous state is also observed for the case with a single star. This can be seen as a bifurcation as the synchronous state appearing in the in phase state and is a consequence of the inside coupling strength $\lambda_{1}$ and $\lambda_{2}$ large enough. On another hand, if we put the inside coupling strength $\lambda_{1}=\lambda_{2}\equiv\lambda$ fixed, and changing the coupling strength $\lambda_0$ between the stars, there is only the first part of synchronization, the abrupt transition to the state with leaf nodes synchronized, as shown in Fig.\ref{fig:9}(b). Hence the two parts of synchronization can be seen as two kinds of mechanics, caused by $\lambda_{1}$ and $\lambda_{2}$, or $\lambda_{0}$, and it is possible to study the two steps of phase transition through controlling $\lambda_{1,2}$ or $\lambda_{0}$ respectively.

For each star, the influence of the other is only a term added to the dynamics of the hub, which is small compared to the total influence of its leaf nodes. Hence, before the transition to the state with leaf nodes synchronized, the state of each of the stars is almost the same with neutral state for a single star network, which is also can be seen in Fig.\ref{fig:9}(b) before the transition. To find the mechanics of the first transition from neutral state to in phase state, we can assume the star 2 is just not affected by star 1 at all, consequently in its neutral state, and focus on how the state of star 1 is affected by it through the unidirectional coupling. As a result, with the increasing of $\lambda_0$, the appearance of the phase transition in star 1 is still observed as shown in Fig.\ref{fig:9}(c).

Furthermore, a more simplified model can be considered, as the limit of neutral state, with $\lambda_{2}=0$ that the star 1 only driven by the hub of star 2. This simplified model reads
\begin{equation}
\begin{aligned}
\dot{\theta}_{h_{1}}&=\omega_{h_{1}}+\lambda\, \sum_{j_{1}=1}^{K_{1}}\sin\,(\theta_{j_{1}}-\theta_{h_{1}})+\lambda_{0}\sin(\theta_{h_{2}}-\theta_{h_{1}}),\\
\dot{\theta}_{j_{1}}&=\omega_{1}+\lambda\,\sin\,(\theta_{h_{1}}-\theta_{j_{1}}), \quad for \; 1\leq j_{1}\leq K_{1},\\
\dot{\theta}_{h_{2}}&=\omega_{h_{2}}.
\end{aligned}
\end{equation}
The drive effect between the star 1 and the hub of star 2 is depend on the coupling strength $\lambda_0$. To check the rational of the new simple model we still set the $\lambda$ among the star 1 unchanged, and reinforce the coupling strength $\lambda_0$ from 0. Amazingly to some extent, the abrupt phase transition to in phase state in star 1 occurs if $\lambda_0$ achieve a critical strength as shown in Fig.\ref{fig:9}(d), just as the original model.
\begin{figure}
  \includegraphics[width=9.5cm,height=7cm]{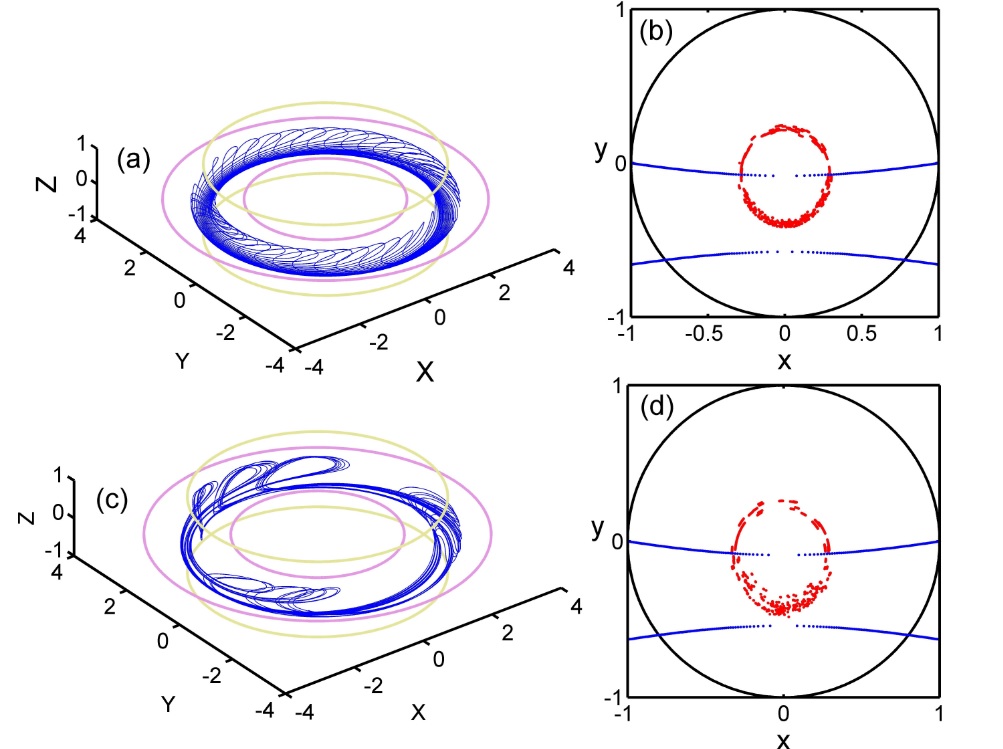}\\
  \caption{(color online) ($a$) and ($c$) The phase space with $\lambda=1.3$, $\lambda_0=0$ or $0.64$ respectively. The blue lines are the trajectories of the simplification model with the transform $X(t)=(3+x(t)*cos(\alpha(t)))$, $Y(t)=(3+x(t)*sin(\alpha(t)))$, $Z(t)=\alpha(t)$, other color lines are the boundary of it. ($b$) and ($d$) The Poincare section of them when $\alpha=\frac{3\pi}{2}$. The red points are the trajectories in this section, the blue lines are $\dot{x}=0$.}\label{fig:10}
\end{figure}

Through the analysis above, it is reasonable to us that the major mechanism that responsible for the first part of phase transition is maintained in this process of simplification. Fortunately, the simplified system is now within our ability to analyze analytically through some method.

To study the mechanism of the phase transition from neutral state to in phase state, we define the order parameter of star 1 as $z_1={r_1}{e^{i\Phi}}=\frac{1}{N}\sum_{i=1}^{N}{e}^{i\phi_i}$ where the phase differences $\phi_i=\theta_i-\theta_{h_1},\alpha=\theta_{h_2}-\theta_{h_1}$ and ${N=K_1}$, with the OA anasatz we can rewrite the model as
\begin{equation}
\begin{aligned}
\dot{z_1}&=i{\triangle\omega}{z_1}-\frac{\lambda}{2}({z_1}^2-1)-\frac{N{\lambda}}{2}({z_1}-{z_1}^*){z_1}-i{\lambda_0}{z_1}\sin{\alpha},\\
\dot{\alpha}&={\triangle\omega}'-\frac{N\lambda}{2i}({z_1}-{z_1}^*){z_1}-{\lambda_0}\sin{\alpha},\\
\end{aligned}
\end{equation}
where ${\triangle\omega}=\omega_1-\omega_h$,${\triangle\omega}'=\omega_{h_2}-\omega_{h_1}$.

By setting $z_1=x+iy$, the system can be described as
\begin{equation}
\begin{aligned}
\dot{x}&=-{\triangle\omega}y-\frac{\lambda}{2}(x^2-y^2-1)+\frac{N\lambda}{2}{y^2}+{\lambda_0}\sin{\alpha}y,\\
\dot{y}&={\triangle\omega}x-{\lambda}xy-N{\lambda}xy-{\lambda_0}\sin{\alpha}x,\\
\dot{\alpha}&={\triangle\omega}'-N{\lambda}y-{\lambda_0}\sin{\alpha},\\
\end{aligned}
\end{equation}
It is worthy to note that this new model is quasi-Hamiltonian also, with the time reversal transformation $R$ which can be resolved into $R=TW$ with $T:t\longmapsto-t$ and $W:(x,y,\alpha)\longmapsto(-x,y,\pm\pi-\alpha)$. Then the trajectories of the system are also time reversible and the volume of phase plane is conserved for reversible trajectories which cross the invariant set $x=0,\alpha=\pm\frac{1}{2}\pi$ more than once. But the system here is three-dimensional, which is more complicated than the two-dimensional one we discussed above. For instance, around the reversible trajectories there can be chaotic trajectories which also has conserved volume, similar as the case of Hamilton system, and the attractors in this system will not be limited by fixed point and limit cycle that limit torus and chaotic attractor are also possible. To fully analyze the system is out of the scope of this paper, we will only focus on the major property of it related to the transition process.
\subsection{States and transitions}
By setting $\dot{x}=0,\dot{y}=0$ and $\dot{\alpha}=0$ when $\lambda_0\neq0$, we have several fixed points in the phase space. However, they are both nonphysical due to the values of $x^2+y^2$ of them are greater than 1 which obey the inner constraints of the order parameter $z$ that the point is allowable to describe the state of system if and only if $x^2+y^2\leq1$. This means that the system can not reach the synchronous state as shown in Fig.\ref{fig:9}(d).

\begin{figure}
  \includegraphics[width=9.5cm,height=7cm]{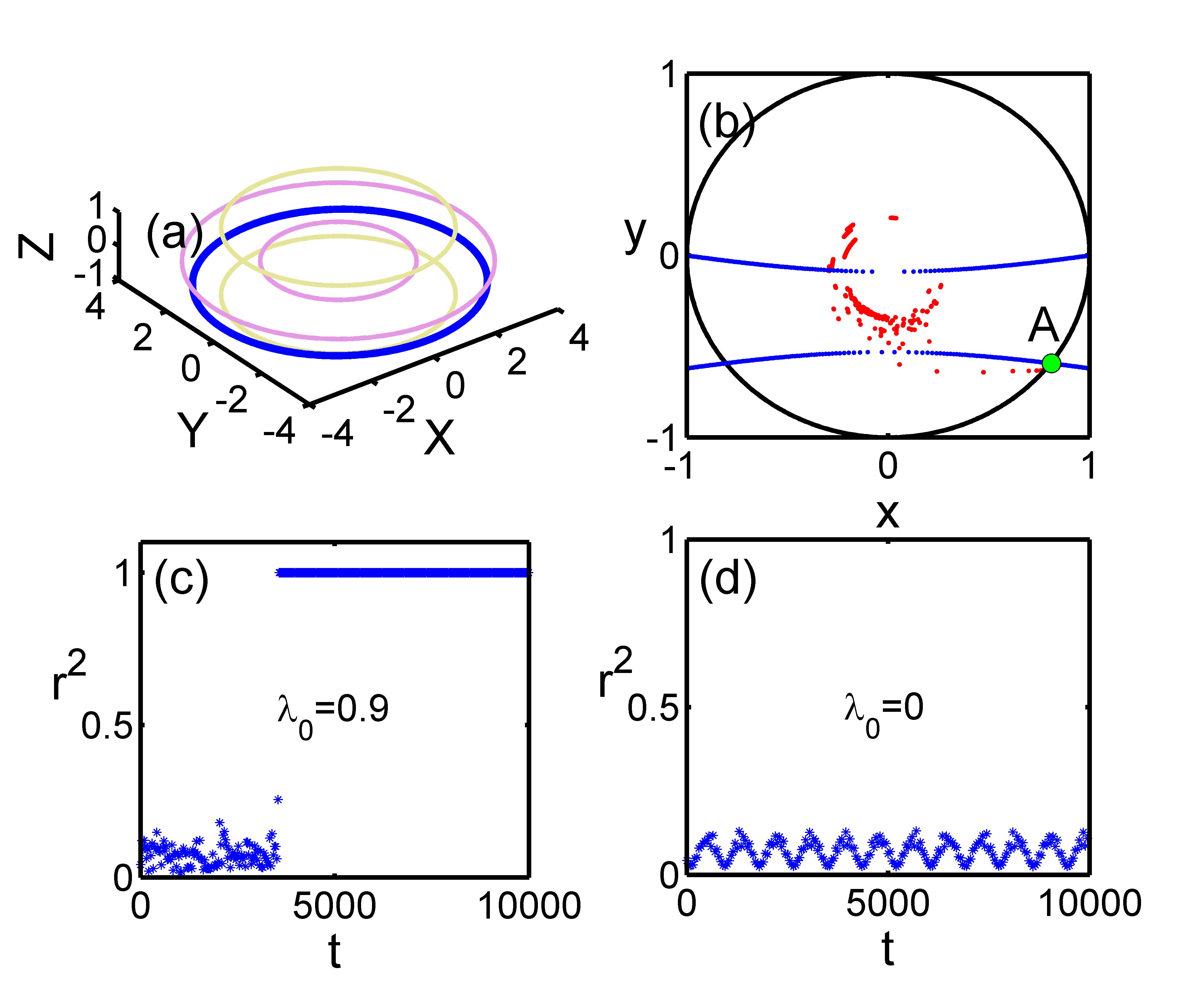}\\
  \caption{(color online) ($a$) The phase space of the simplification model with $\lambda=1.3, \lambda_0=0.9$. The blue lines are the trajectories with the transform $X(t)=(3+x(t)*cos(\alpha(t)))$, $Y(t)=(3+x(t)*sin(\alpha(t)))$, $Z(t)=\alpha(t)$, other color lines are the boundary of it. ($b$) The Poincare section with $\alpha=\frac{3\pi}{2}$ of the phase space, the red points are the trajectories in this section, the blue lines are $\dot{x}=0$, the fixed point A is the intersection of down blue line $\dot{x}=0$ and the unit cycle. ($c$)-($d$) The local order parameter of star 1 against the time with different $\lambda_0$.}\label{fig:11}
\end{figure}

In another hand, for this system, the state with all the leaf nodes synchronized is no doubt a limit circle too, for only one degree of freedom left, shown in Fig.\ref{fig:11}(a). As for the neutral stable state, when $\lambda_{0}=0$ it is either the periodic trajectory or quasi-periodic torus, shown in Fig.\ref{fig:10}(a), and when $\lambda_{0}>0$ it is disturbed by the coupling between stars which may lead the state to be chaotic as shown in in Fig.\ref{fig:10}(c). Hence the transition from the neutral state to in phase state in this system is described by the transition from the chaotic or quasi-perotic torus to a limit circle.

Even though the system has many dynamical structures as chaotic attractors and so on, what we focus on is only the transition from these two kinds of state which is the major factor to the problem. As to the case with two dimension dynamics, the abrupt transition and hysteresis can be described by the separation line for the two states in the coexistence region. And to a system with time-reversal symmetry the separation line is right the nullclines which prevent the trajectory crossing the invariant set twice, as $\dot x=0$ for a single star. For the three-dimensional system here, the cases are more complicated, but the properties related to the time-reversal symmetry is quite similar. The separation for the two states, volume conserved trajectory and limit circle, are the surface with $\dot x=0$. When $\lambda_{0}=0$, it reduce to the case for a single star. The surface reads
\begin{equation}
-\bigtriangleup\omega{y}-\frac{\lambda}{2}(x^2-y^2-1)+N\lambda{y^2}+{\lambda_0}{\sin\alpha}y=0
\end{equation}
the surface is wavy and it will be the peak or bottom when $\alpha=\frac{\pi}{2}$ or $\frac{3\pi}{2}$, which means that the largest distance between a point in the surface and a point in the tours around center is with $\alpha=\frac{\pi}{2}$, and the smallest with $\frac{3\pi}{2}$. It is coincide with symmetry analysis that the invariant sets are exactly with $\alpha=\frac{\pi}{2}$ or $\frac{3\pi}{2}$.

For case for a single star, the neutral state depends on the initial state and if its trajectory hit the separation line it will evolve to synchronous state, otherwise it will stay as the neutral state. The position of separation line depends on parameters which gives the basion of attraction of synchronous sate and also the critical coupling strength for fixed initial condition. For the case with three dimension, it is quite similar, except that the synchronous state for a single star is now the limit circle and the separation line is the separation surface for the system.

For the wavy structure of the separation surface, we can only consider the bottom where the surface is closest to the torus. For $\frac{3\pi}{2}$, the variable of $x^2$ can be formed by $y$ that
\begin{equation}
x^2=y^2+1+\frac{2}{\lambda}(-\lambda_0-\bigtriangleup\omega)y+2Ny^2
\end{equation}
then the range of $y$ is $[-1,\frac{-b-\sqrt{b^2-4(2N+1)}}{4N+2}]$ with $b=\frac{2}{\lambda}(-\lambda_0-\bigtriangleup\omega)$. Together with range of $y$ and the relation with $x$ and $y$ we could get the distance between the separation surface and the torus with order parameter $r_{0}$
\begin{equation}
\begin{aligned}
d=&r_{min}-r_{0}\\
&\frac{2[b^2+b\sqrt{b^2-4(2N+1)}-(4N+2)(4N+4)]}{(4N+2)^2}+1-r_{0}
\end{aligned}
\end{equation}
If $d\leq0$ the separation surface and the torus have intersection and if $d>0$ the two surfaces are separated. The critical coupling strength for a fixed initial condition is the case with $d=0$. If the number of N is large enough, $r_{min}$ approximate equal to $\frac{b}{2N}$, which means that when the phase transition occurs, the order parameter $r$ has a linear relation with the critical coupling strength $\lambda_{0c}$. Increasing $\lambda_{0}$, $r_{min}$ decease and more torus will intersect with separation surface and more trajectories will evolve to the limit circle.

As an example and illustration, the process is shown in Fig.\ref{fig:10} and Fig.\ref{fig:11}. When $\lambda_0=0$, the star 2 has no influence on the star 1, as in Fig.\ref{fig:10}(a) and (b), and the star 1 is in neutral stable state which we have discussed above with the trajectories in the phase space regular and periodical. When $\lambda_0=0.7$, the star 2 has greater influence on the star 1, the trajectories in the phase space become chaotic and irregular, from the view of Poincare section, the red points jitter and expand greatly. Meanwhile, the separation line, as the down blue one in the figure, upwards all the time with the increasing of $\lambda_0$, and the volume of the basin of attraction of in phase state is increasing. When the coupling strength $\lambda_0$ between hubs is big enough, the trajectories will hit the separation line and evolve to the limit cycle which corresponds to the in phase state, as shown in Fig.\ref{fig:11}, when $\lambda_0=0.9$. The phase transition occurs now, and all the leaf nodes synchronize in this time.

\begin{figure}
  \includegraphics[width=9.5cm,height=4cm]{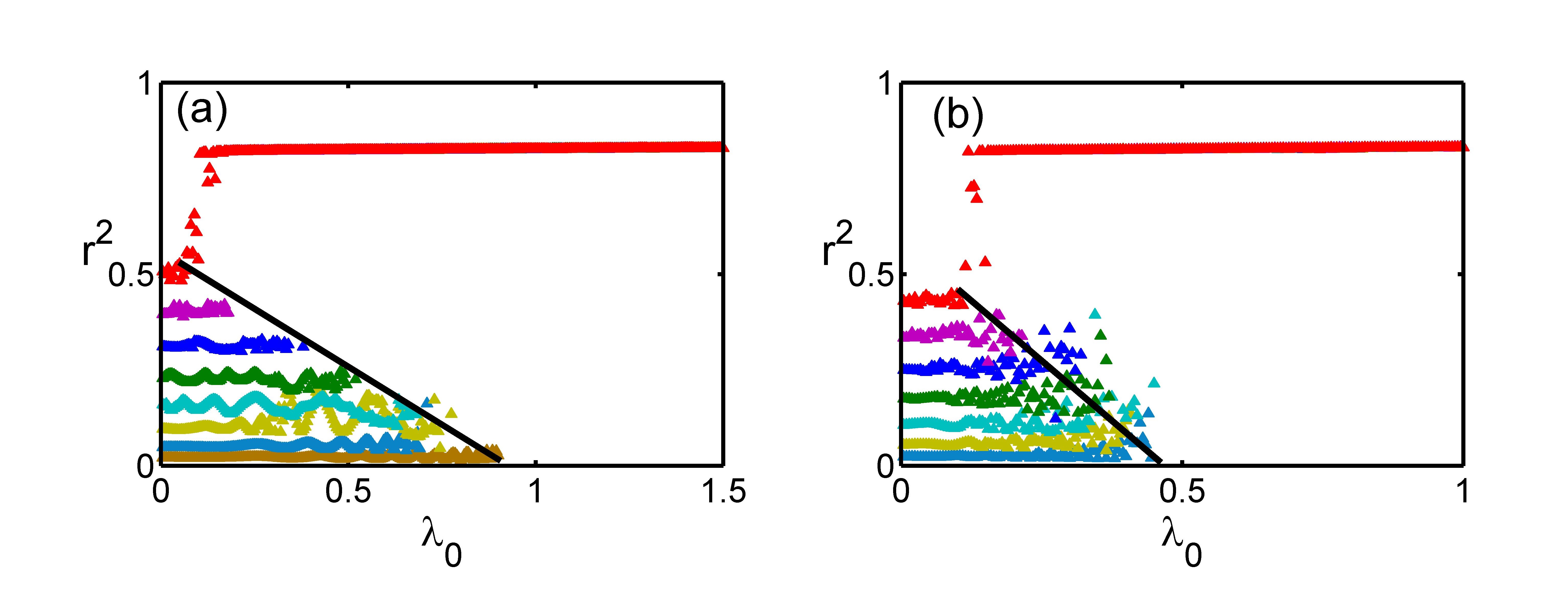}\\
  \caption{(color online) The order parameter of star 1 against the coupling strength $\lambda_0$ with different initial states ($a$) with $K_1=10,K_2=10, \lambda_1=\lambda_2=\lambda=0.5$, unidirectional coupled by the star 2, ($b$) with $K_1=10,K_2=8, \lambda_1=\lambda_2=\lambda=0.8$, bidirectional coupled with the star 2.}\label{fig:12}
\end{figure}

Compared with the case with a single star, this dynamical process is quite similar, except three points following. First, the fixed point as the synchronous state is related to the limit cycle solution as the in phase state. Second, the neutral state is now chaotic and inflates because of chaos. And third, the coupling between the stars will enlarge the basion of attraction of the in phase state, reducing the region for neutral state. Because of the three factors, we have the mechanism of the first step of transition from neutral state to in phase state, which can be checked by the linear relation between $\lambda_{0c}$ and initial condition $r_{0}$, shown in Fig.\ref{fig:12} for both the unidirectional and bidirectional coupled star networks.

Regardless of the coupling strength between the two star networks either unidirectional or bidirectional, the linear relation between the initial order parameter and the critical coupling strength $\lambda_{0c}$ are both observed. The linear relations in those models are in agreement with the result we get, supporting the analysis above and the mechanism we get from the simplified model.

\section{Conclusion}
In this paper, the star network and star connected network are used to study the main character of the heterogenous network. According to the approximated OA ansatz method, we propose the ensemble order parameter equation. Based on this equation, the high-dimensional phase space can be reduced to a low-dimensional order parameter space without additional approximation, then the essential dynamical mechanism can be analyzed analytically. In the star network, four characteristic states are found by the ensemble order parameter equation and the characters and stable regions of them in the phase diagram are determined by the time reversibility analysis and linear stability analysis. Under the action of phase shift and coupling strength, extensive phase transitions among those states are found. Three processes of synchronization from splay state, in phase state and neutral state to synchronous state respectively, the de-synchronization from synchronous state to splay state and a group of hybrid transitions in the synchronous state, the splay state and from in phase state to splay state or the opposite. The processes and mechanisms of them are discussed by means of the plane of ensemble order parameter, with planar dynamical of it, all the properties of phase transitions can be determined analytically. In the star-connected network, we discover the phenomena of two-step phase transition. Through the simplification, we get the mechanism of this transition through the similar method as the case for a single star network, namely ensemble OA method. A linear relation between initial order parameter and critical coupling strength between stars are get and checked.

\end{document}